\documentclass[aps,prb,reprint, superscriptaddress,twocolumn,showpacs]{revtex4-1}

\usepackage{graphicx, amsmath} \usepackage[colorlinks=true,allcolors=blue]{hyperref}

\newcommand{\ZSS}{ZrSiS}

\begin{document}



\title{Anisotropic Berry phase in the Dirac nodal-line semimetal ZrSiS: The effect of spin-orbit coupling}



\author{Yusen Yang} 
\affiliation{Key Laboratory of Artificial Structures and Quantum Control, and Shanghai Center for Complex Physics, School of Physics and Astronomy, Shanghai Jiao Tong University, Shanghai 200240, China} 

\author{Hui Xing} 
\email{huixing@sjtu.edu.cn}
\affiliation{Key Laboratory of Artificial Structures and Quantum Control, and Shanghai Center for Complex Physics, School of Physics and Astronomy, Shanghai Jiao Tong University, Shanghai 200240, China} 

\author{Guoxiong Tang } 
\affiliation{Key Laboratory of Artificial Structures and Quantum Control, and Shanghai Center for Complex Physics, School of Physics and Astronomy, Shanghai Jiao Tong University, Shanghai 200240, China}

\author{Chenqiang Hua} 
\affiliation{Zhejiang Province Key Laboratory of Quantum Technology and Devices, Department of Physics, Zhejiang University, Hangzhou 310027, China}

\author{Chao Yao} 
\affiliation{Key Laboratory of Artificial Structures and Quantum Control, and Shanghai Center for Complex Physics, School of Physics and Astronomy, Shanghai Jiao Tong University, Shanghai 200240, China}

\author{Xiaoxian Yan} 
\affiliation{Key Laboratory of Artificial Structures and Quantum Control, and Shanghai Center for Complex Physics, School of Physics and Astronomy, Shanghai Jiao Tong University, Shanghai 200240, China}

\author{Yunhao Lu} 
\affiliation{Zhejiang Province Key Laboratory of Quantum Technology and Devices, Department of Physics, Zhejiang University, Hangzhou 310027, China}

\author{Jin Hu}
\affiliation{Department of Physics, University of Arkansas, Fayetteville, Arkansas 72701, USA}

\author{Zhiqiang Mao}
\affiliation{Department of Physics and Materials Research Institute, Pennsylvania State University, University Park, Pennsylvania 16802, USA}

\author{Ying Liu} 
\email{yxl15@psu.edu}
\affiliation{Department of Physics and Materials Research Institute, Pennsylvania State University, University Park, Pennsylvania 16802, USA} 

\date{\today}

\begin{abstract}
The topological nodal-line semimetals (NLSMs) possess a loop of Dirac nodes in the $k$ space with linear dispersion, different from the point nodes in Dirac/Weyl semimetals. While the quantum transport associated with the topologically nontrivial Dirac fermions has been investigated extensively, features uniquely associated with the extended nodal lines remain to be demonstrated. Here, we investigate the quantum oscillations (QOs) in the nodal-line semimetal \ZSS\/, with the electron transport along the $c$ axis, and magnetic field rotating in the $ab$ plane. The extremal orbits identified through the field orientation dependence of the QOs interlock with the nodal line, leading to a nonzero Berry phase. Most importantly, the Berry phase shows a significant dependence on the magnetic field orientation, which we argue to be due to the finite spin-orbit coupling gap. Our results demonstrate the importance of the spin-orbit coupling and the nodal-line dispersion in understanding the quantum transport of NLSMs.

\end{abstract} 

\maketitle

\section{INTRODUCTION}

Topological semimetals feature conduction and valence band crossing \cite{Z.K.Liu2014, M.Neupane2014, C.Fang2015}, a topologically distinct property, which garnered extensive attention due to the low energy excitation resembling that of relativistic particles \cite{S.M.Young2012, S.M.Young2015}. A large body of exotic behavior useful for quantum information technology, including high magnetoresistance (MR), high carrier mobilities, and chiral anomaly, have been explored \cite{J.Hu2019}. Protected by the crystalline or the time-reversal symmetry \cite{A.A.Burkov2016}, the band crossing in topological semimetals is stable, which has led to distinct topology-related features, particularly the presence of nonzero Berry phase associated with the nodes in the $k$ space \cite{A.A.Taskin2011,S.L.Huang2017}. 

The band crossing in the $k$ space can be Dirac/Weyl nodes occurring at discrete points, or nodal lines which consist of open lines or closed rings \cite{C.Fang2016,A.Bernevig2018,N.P.Armitage2018}. Linear dispersion near the nodes was confirmed by band structure calculations and measurements. In particular, electrical quantum transport measurements played an indispensable role in the study of topological semimetal for its ability of quantifying the Berry phase \cite{J.Hu2019}. The topological signature of Dirac/Weyl nodes in the quantum transport, due partly to the discrete nodal points which render an ‘isotropic’ Berry curvature, has been well characterized. However, for the nodal-line semimetals (NLSMs), the nodal points at different $k$ values do not locate at the same energy. This dispersion of the nodal lines themselves adds complexity, such as the correlation effect found in ZrSiSe \cite{Y.M. Shao2020}. To date, the physical consequences of the nodal line dispersion in quantum oscillation measurements have not been demonstrated.

One representative topological NLSM, ZrSiS, attracted extensive attention. \ZSS\ possesses a lattice structure with square nets of Si, which turned out to be described by the square-net model proposed by Young and Kane \cite{S.M.Young2015}. One important consequence is that \ZSS\ hosts two types of Dirac nodes: (i) Dirac points protected by the nonsymmorphic symmetry and (ii) Dirac nodal lines protected by the inversion symmetry and time-reversal symmetry \cite{L.M. Schoop2016}. The two types of Dirac line nodes is different in its response to spin-orbit coupling (SOC): the former is immune to SOC, while the latter will open a gap. However, the former is far below the Fermi energy, therefore the Fermi surface (FS) in \ZSS\ is purely consisted of linearly dispersed bands close to the nodal lines, for which the effect of SOC is non-negligible. The linear band dispersions persist up to 2 eV from the Fermi level \cite{M.Neupane2016}. As a result, the electron transport is dominated by the Dirac fermions, providing an ideal system to explore the effect of nodal lines on the transport properties.

The FS of \ZSS\ has been characterized by a number of experiments, including ARPES \cite{M.Neupane2016,B.B.Fu2019,C.Chen2017}, de Haas-van Alphen (dHvA) oscillations \cite{J.Hu2017}, Shubnikov–de Haas quantum (SdH) oscillations \cite{R.Singha2017,M.N. Ali2016} and thermoelectric quantum oscillations \cite{M.Matusiak2017}. The FS consists of an electronlike and a holelike band, forming a diamond shape, which fully encloses the nodal line. The Berry curvature associated with the nodal line was found in earlier quantum oscillation measurements,most of which adopted a configuration with the electron transport in the $ab$ plane and a magnetic field parallel to the $c$ axis. A phase shift in the oscillations was identified clearly as a consequence of the topology of the band structure. Nevertheless, the reported values of phase shift bear a significant variation. Little insight on the effect of nodal-line dispersion and SOC was obtained due to the limited measurement configurations. Here, we performed measurements of SdH oscillations on \ZSS\ single crystals with the electron transport along the $c$ axis and a rotating magnetic field in the $ab$ plane, where the nodal line manifested a significant dispersion. A systematic variation of the SdH oscillations was observed upon the change of the azimuthal angle. However, the Berry phase deduced consistently from both the Landau fan diagram and the Lifshitz-Kosevich formula shows significant variation. The configuration between the FS and the nodal lines requires a nontrivial Berry phase for all azimuthal angles. We argue that the change in the Berry phase found in our experiments is due to the effect of SOC on the dispersed nodal lines, an effect often overlooked in previous experiments. 

\begin{figure}[t] 
	\centering \includegraphics[width=1\columnwidth]{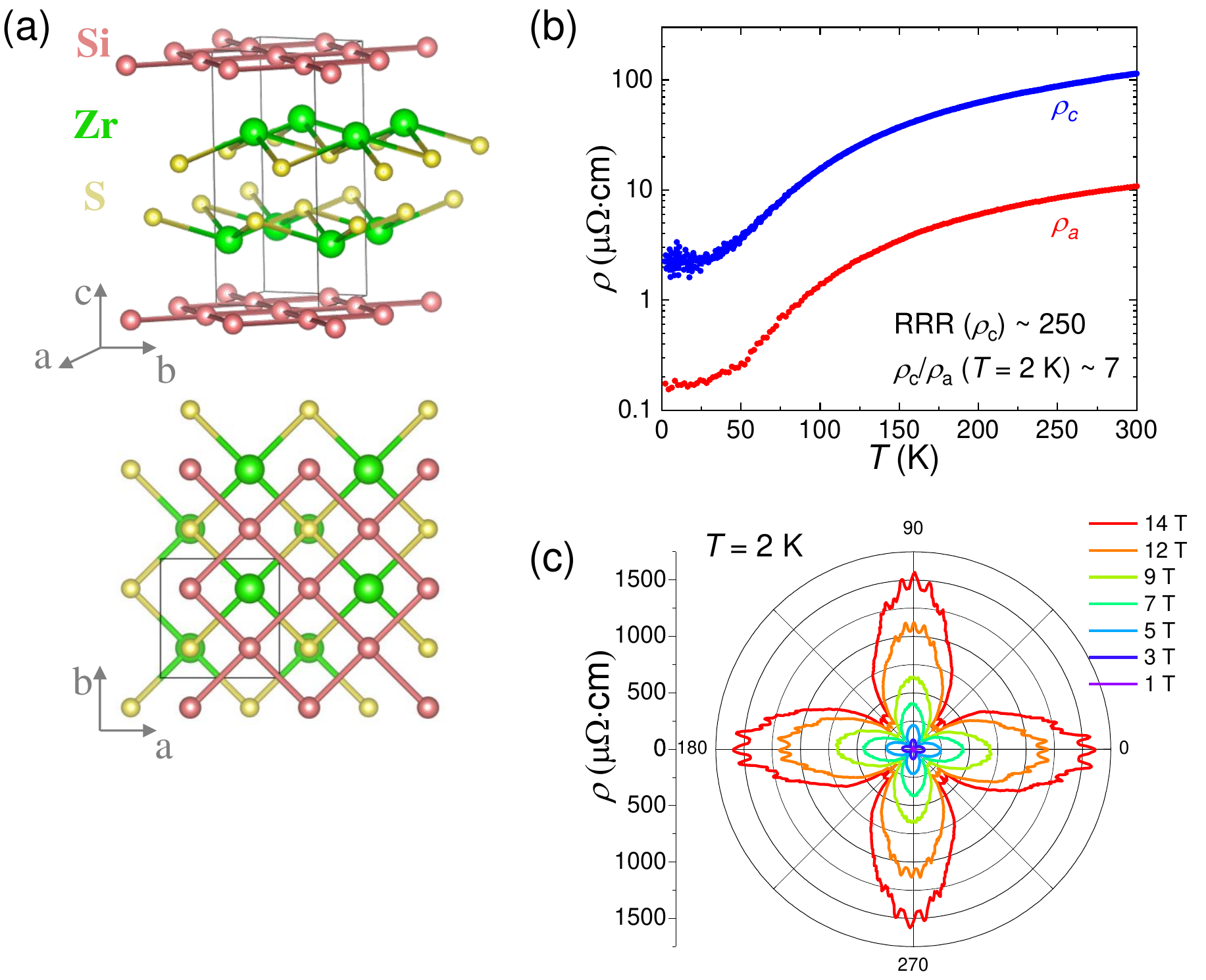} 
	\caption{(a) Tetragonal crystal structure of \ZSS\ consisting of quintuple layers of S-Zr-Si-Zr-S. The neighboring S layers are weakly bonded which forms natural cleavage planes. (b) Zero-field in-plane ($\rho_a$) and out-of-plane ($\rho_c$) resistivity. Resistivities in both directions show metallic behavior with anisotropy $\rho_{c} / \rho_{a}$(T = 2 K) $\sim 7$ and excellent crystal quality with a RRR $\sim$ 250 in $\rho_c$. (c) Azimuthal angle dependent MR with the current applied along the $c$ axis and magnetic field rotating in the $ab$ plane at 2 K. The MR shows apparent fourfold symmetry. SdH oscillations were observed in a large magnetic field around the principal axes.} \end{figure}

\section{EXPERIMENTAL METHOD}

Single crystals of \ZSS\ were grown by the chemical vapor transport method. \ZSS\ crystals were characterized by powder x-ray diffraction (XRD) to confirm purity and single-crystal XRD with the Laue method to determine crystal orientation. All crystals were polished to be a regular shape which enables in-plane and out-of-plane transport measurements. The sample quality is further confirmed by a large residual resistance ratio of ~250. In-plane and out-of-plane resistance were measured with current applied along the [100] and [001] direction respectively by using the standard four terminal method in a Quantum Design physical property measurement system (PPMS) with 14 T magnet and ac transport and resistivity options. The angle dependent MR was measured with a rotator for controlling the angle between the magnetic field and crystal axis.

\section{RESULTS AND DISCUSSION}

The tetragonal crystal structure of \ZSS\ is shown in Fig. 1(a), which consists of quintuple layers of S-Zr-Si-Zr-S. One particular feature of the lattice is that Si forms square networks, as can be seen from the top view of the lattice. The crystal possesses a glide-mirror symmetry with respect to the Si layer. The neighboring S layers are weakly bonded which form natural cleavage planes \cite{R.Sankar2017,C.C. Su2018}. Both the in-plane (resistivity along the $a$ axis, $\rho_a$) and out-of-plane (resistivity along the $c$ axis, $\rho_c$) resistivities at zero magnetic field show metallic behavior, as seen in Fig. 1(b), with the resistivity anisotropy $\rho_c$/$\rho_a$ around 7 at 2 K, consistent with earlier reports \cite{K.R. Shirer2019}. The residual resistance ratio (RRR) along the $c$ axis is found to be up to 250, confirming the high crystal quality. The similar temperature dependence of both in- and out-of-plane MR signals a similar dominant transport mechanism for both directions. 

The anisotropy of the system was also investigated with the current applied along the $c$ axis with the magnetic field rotating in the $ab$ plane. This measurement configuration ensures the orthogonal relative orientation between the current and the magnetic field for all azimuthal angles. In Fig. 1(c), the azimuthal angle dependent values of MR for various magnetic fields at 2 K are shown. An apparent fourfold symmetry is evident for all magnetic fields. \ZSS\ possesses maximal resistivity for $H$ parallel to the principal in-plane axes and minimal resistivity for $H$ along the bisector in-plane axes. A close examination of the data reveals a small and complex pattern near the bisector directions, which is similar to earlier measurements \cite{K.R. Shirer2019,M.Novak2019}. The exact nature of this behavior is not yet clear, although the detailed FS morphology may be responsible. In addition, SdH oscillations were found to emerge with the increasing field, manifested as small peaks seen around the principal axes. The angular dependence of SdH oscillations is determined by the anisotropy of the FS in the in-plane direction.

\begin{figure*}[t] 
	\centering \includegraphics[width=2\columnwidth]{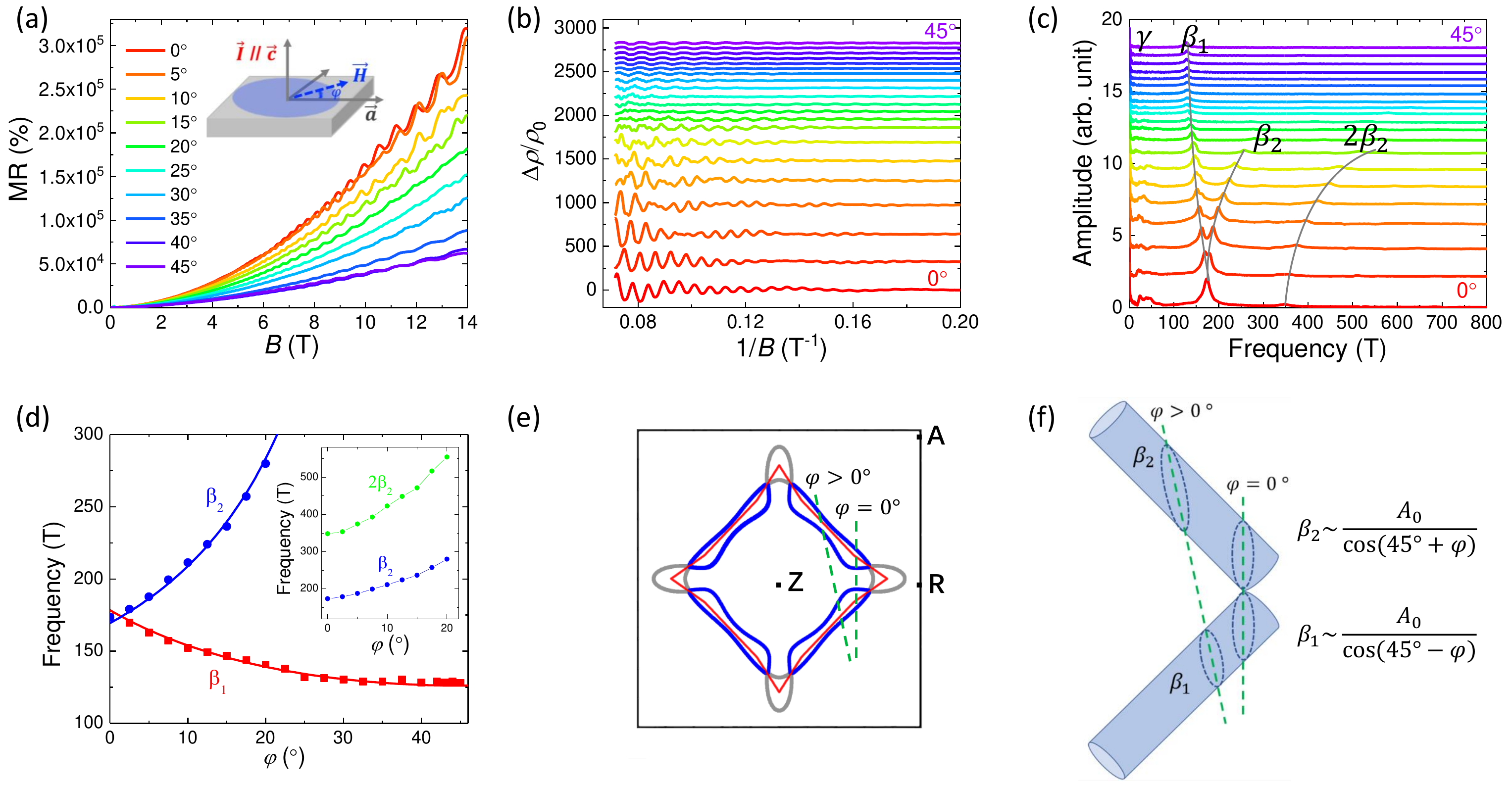} 
	\caption{(a) Magnetoresistance with current applied along the $c$ axis and magnetic field at various azimuth angle $\varphi$. The magnitude of the magnetoresistance reaches as high as $3.2\times 10^5$ percent at 14 T and 2 K. The inset shows the schematic of the measurement configuration. (b) Extracted oscillatory part in magnetoresistance at various angles by subtracting the nonoscillatory background. (c) Fast Fourier Transformation (FFT) spectra of the SdH oscillations at various angles. The grey solid lines are guides to illustrate the frequency evolution. (d) Angle dependence of the resolvable frequencies $\beta_1$, $\beta_2$ and 2$\beta_2$. The red and blue solid lines are fittings using the simplified model in (f). The simplified cylinder model fits the $\beta_1$ and $\beta_2$ branches very well, as shown by the solid lines. (e) Schematic FS in the Z–R–A plane, adopted form Ref. \cite{S.Pezzini2018}. The FS consists of an electronlike dog-bone (indicated by the blue line) and a holelike petal part (grey line). (f) A simplified model of the FS in (e). The electronlike dog-bone bands are approximated by perpendicular cylinders. The corresponding extremal cross sections are indicated by the blue dash circles, with the angle dependence of $\beta_{1} \sim \frac{A_{0}}{\cos \left(45^{\circ}-\varphi\right)}$ and $\beta_{2} \sim \frac{A_{0}}{\cos \left(45^{\circ}+\varphi\right)}$, respectively.} \end{figure*}

The MR for the same measurement configuration ($I // c$, $H // ab$ plane) is shown in Fig. 2(a), which exhibits nearly quadratic field dependences at various azimuthal angles $\varphi$. The subquadratic field dependence is commonly observed in topological semimetals \cite{J.Hu2019}. The MR magnitude decreases monotonically with $\varphi$ increased from 0$^{\circ}$ to 45$^{\circ}$, again owing to the FS anisotropy. Most notably, the MR reaches as high as $3.2\times 10^5 \% $  at 14 T and 2 K for $\varphi$ = 0$^{\circ}$ with nonsaturating trend. Similar results were obtained in previous studies and attributed to the close-to-compensate electron and hole carriers \cite{J.A.Voerman2019}.

The prominent oscillatory part in the MR can be extracted by subtracting the nonoscillatory background. The corresponding fast Fourier transformation (FFT) spectra are shown in Fig. 2(c). There are three major features in the FFT spectra:  (i) The peak with the lowest frequency $\gamma$ around 30 T, which is weakly angle dependent, is close to that found in the intra-plane transport \cite{M.N. Ali2016}. (ii) The peak at 174 T for $\varphi$ = 0$^{\circ}$ is seen to bifurcate into two branches with opposite angular dependences (denoted as $\beta_1$ and $\beta_2$, respectively), and marked with grey solid lines. (iii) The peak with the higher frequency is noted as the double-frequency peak of  $\beta_2$. We note that an additional small shoulder peak appeared near $\beta_1$, which could be the effect of Zeeman splitting \cite{J.A.Voerman2019}.  In addition, several small peaks with very low frequencies appear for small $\varphi$, which are likely due to the area near the crossing point of the nodal lines, as proposed recently \cite{G.P.Mikitik2020}.

\begin{figure*}[t] \centering \includegraphics[width=2.0\columnwidth]{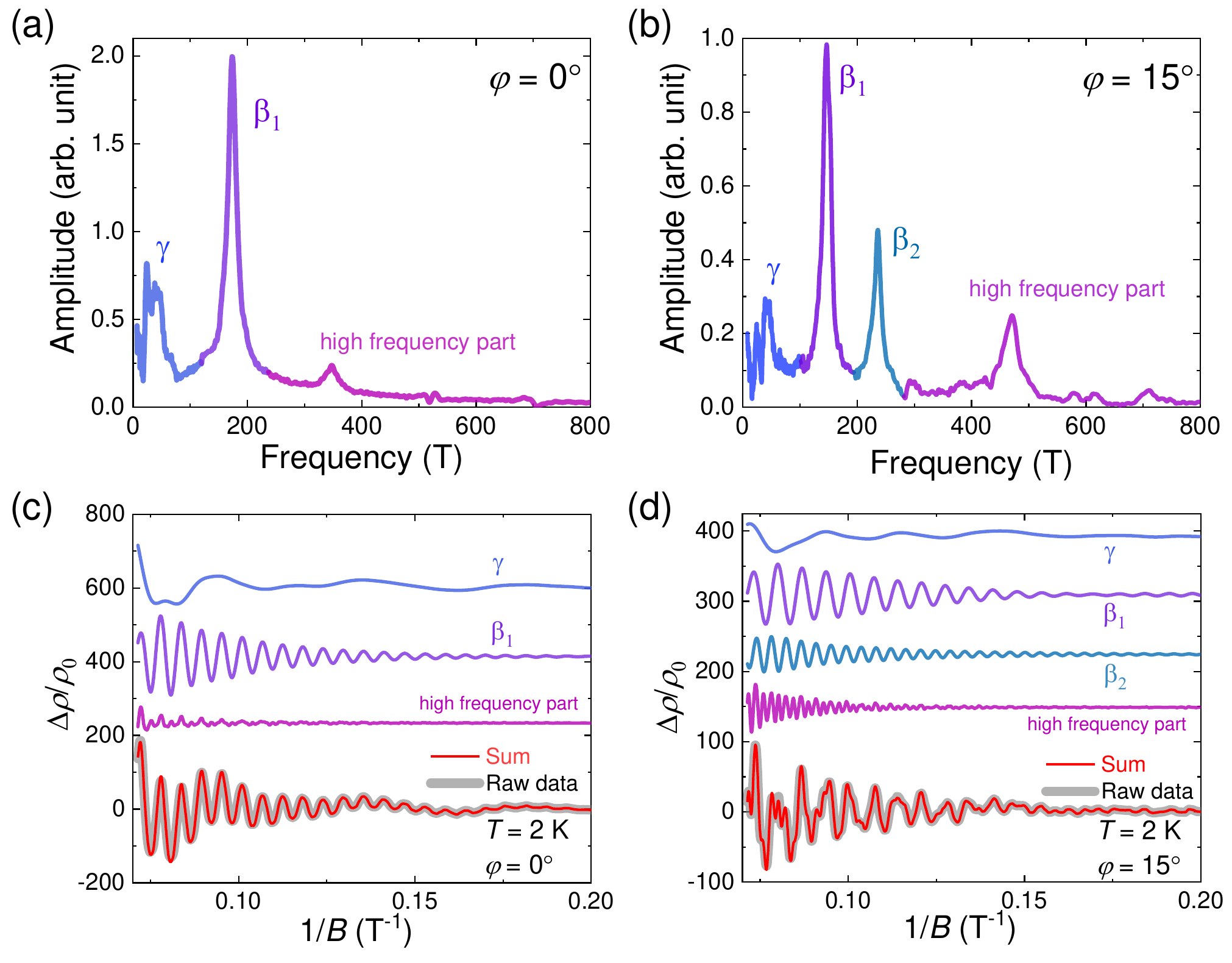} 
	\caption{[(a), (b)] FFT spectra of the quantum oscillations with magnetic field at $\varphi$ = 0$^{\circ}$ and 15$^{\circ}$ for $T$ = 2 K. [(c), (d)] The extracted single-frequency oscillatory parts at $\varphi$ = 0$^{\circ}$ and 15$^{\circ}$. The sum of all the extracted oscillatory components matches with the raw data.} \end{figure*}

To obtain a quantitative understanding of the oscillation frequencies described above, additional information on the FS is needed. However, the nodal-line semimetallic nature of \ZSS\ makes its FS structure very sensitive to the exact position of the Fermi level, which is partly the reason for the apparently different FS reported in earlier DFT calculations \cite{B.B.Fu2019,S.Pezzini2018,A.N.Rudenko2020}. Among the resolvable peaks ($\gamma$, $\beta_1$ and $\beta_2$) summarized in Fig. 2(c), we will focus on the two better-resolved peaks $\beta_1$ and $\beta_2$ whose azimuthal angle dependence is shown in Fig. 2(d). The two peaks emerge from the same 174 T frequency at $\varphi$ = 0$^{\circ}$. With increasing $\varphi$, $\beta_1$ deceases progressively, while $\beta_2$ increases. These features provide an important clue for the origin of these two peaks. The most likely FS that corresponds to the carrier density in our crystals is shown in the schematic in Fig. 2(e), which is the FS of \ZSS\ at $k_z=\frac{\pi}{c}$ reproduced from Ref. \cite{S.Pezzini2018}. The FS consists of a dog-bone electronlike and a petal holelike pocket. If we approximate this FS by two cylinders as shown in Fig. 2(f), two extremal cross sections exist with opposite $\varphi$ dependences, $\beta_{1} \sim \frac{A_{0}}{\cos \left(45^{\circ}-\varphi\right)}$ and $\beta_{2} \sim \frac{A_{0}}{\cos \left(45^{\circ}+\varphi\right)}$, respectively, where $A_0$ is the base area of the cylinder. It is evident that upon increasing azimuth angle $\varphi$, the cross-sectional areas for $\beta_1$ and $\beta_2$ start off with equal size, followed by one decreasing and the other increasing. We found that the angle dependences of the frequencies can fit the experimental data very well, as shown by the red and blue solid lines in Fig. 2(d). The resultant fitting parameter $A_0$ for the $\beta_1$ and $\beta_2$ branches are 126 T and 120 T, respectively, reasonably close to one another. The quality of the fitting suggests that quantum oscillations of $\beta_1$ and $\beta_2$ correspond to the extremal orbits in the electronlike bands as depicted in Fig. 2(e). 

The well-defined QOs permits a quantitative analysis of its underlying physical parameters. For that we need to extract the oscillatory components associated with discrete frequencies. As shown in Figs. 3(a) and 3(b), the raw oscillation data were treated between the cutoff frequencies marked with different colors. The extracted single-frequency oscillatory parts are plotted separately in Figs. 3(c)and 3(d). The reliability of such extraction is seen in the match between the sum of the three extracted oscillations and the raw data. This procedure is applied to MR oscillations at all azimuthal angles. 

\begin{figure*}[t] \centering \includegraphics[width=2\columnwidth]{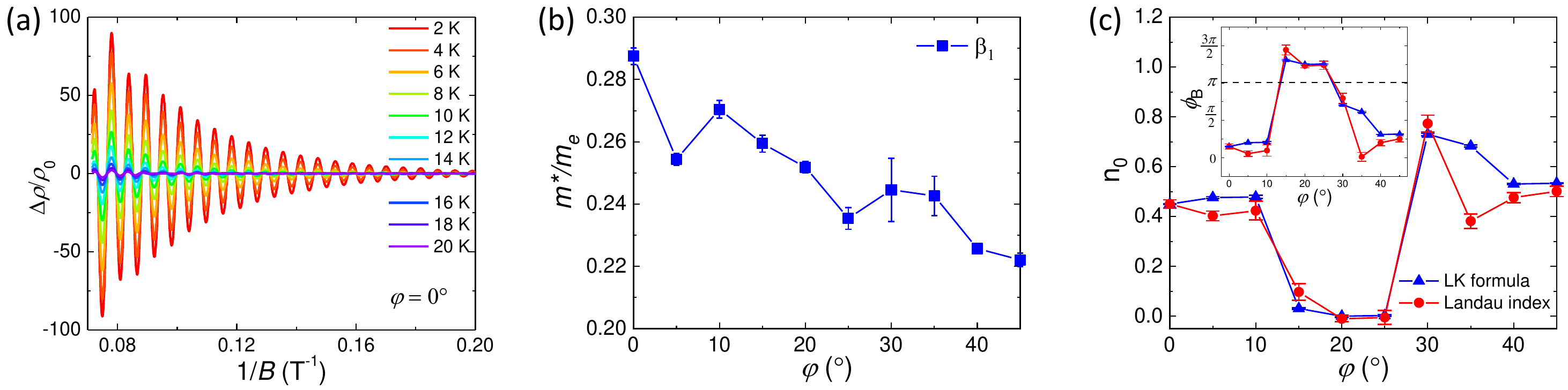} 
	\caption{(a) The $\beta_1$ oscillatory component of the magnetoresistance at various temperature from 2 K to 20 K at $\varphi$ = 0$^{\circ}$. (b) The angle dependence of effective mass fitted from temperature dependence of the oscillatory amplitude. (c) The angle dependence of total phase extracted by fitting with LK formula and Landau fan diagram, respectively (see text). The inset shows the corresponding Berry phase indicating its nonzero value for all azimuthal angles. Error bars count the uncertainty in fitting the single-frequency oscillation only.} \end{figure*}

As we pointed out above, the hallmark of a topological semimetal in transport is a distinct Berry phase shift. The Lifshitz-Kosevich (LK) formula describes the SdH oscillation, $\Delta\rho_{xx}\propto R_{S} R_{T} R_{D} \cos \left[2 \pi\left(\frac{F}{B}+n_{0}\right)\right]$, where $R_{S}=\cos \left(\frac{\pi g^{*} m^{*}}{2 m_{e}\left(1+\lambda_{e-p h}\right)}\right)$, $R_{T}=\frac{2 \pi^{2} k_{B} T m^{*} / \hbar e B}{\sinh \left(2 \pi^{2} k_{B} T m^{*} / \hbar e B\right)}$ and $R_{D}=e^{-2 \pi^{2} k_{B} T_{D} m^{*} / \hbar e B}$ are spin, thermal and Dingle damping terms, respectively, and $g^{*}$ is the effective $g$ factor, $m^{*}$ the effective mass of the quasiparticle, $m_e$ the electron mass, $\lambda_{e-ph}$ the electron-phonon coupling strength, $k_B$ the Boltzmann constant, and $T_D$ the dingle temperature strength. In particular, $n_{0}=-\frac{1}{2}+\beta \pm \delta_{L K}$ is the total phase, with Berry phase $\phi_B=2\pi\beta$ and phase shift $\delta_{LK}$. $\delta_{LK}$ is related to the curvature of the FS with 0 for two dimensions and $\pm \frac{1}{8}$ for three dimensions. Fig. 4(a) presents the temperature dependence of the oscillatory amplitude with field at $\varphi$ = 0$^{\circ}$. The effective mass $m^\ast$ can be extracted through the fitting of the temperature dependence of the oscillatory amplitude using the thermal damping term $R_T$ in the LK formula. The resultant effective mass is about  0.29 electron mass, with a small angular dependent variation in the range of 0.22 to 0.29 as shown in Fig. 4(b). Most importantly, the total phase $n_0$ obtained from the LK fitting shows significant angle dependence, as shown in Fig. 4(c). The reliability of the extracted value of $n_0$ is further verified by treating the same quantum oscillation data using the Landau fan diagram. Values of $n_0$ obtained from both methods are consistent.

The total phase $n_0$ is determined by the dimensionality of the FS and the possible Berry phase, as $n_0=-\frac{1}{2}+\frac{\phi_B}{2\pi}+\delta_{3D}$. The cyclotron orbit we have been focused on is the maximum cross section in the 3D electronlike band, so that $\delta_{3D}=-\frac{1}{8}$ \cite{C.Q.Li2018}. Therefore, we obtain the corresponding Berry phase $\phi_B=2\pi(n_0+\frac{5}{8})$, as shown in the inset of Fig. 4(c). It can be seen that $\phi_B$ is nonzero for all azimuthal angles. It is close to the ideal Berry phase of $\pi$ for $15^{\circ} \leq \theta \leq 25^{\circ}$, and is close to zero but remains finite for the rest field orientations. The angle dependence of the Berry phase is intriguing. While it is reasonable to see a nonzero Berry phase in a NLSMs, a close examination of the magnetic field orientation with respect to the FS (Fig. 2(e)) shows that the cyclotron orbit interlocks the nodal line for all the field angle in our measurements. Therefore, it is expected to show nonzero Berry phase for all the investigated azimuthal angles. However, the significant variation of the Berry phase requires further interpretation.

It has been noted both theoretically and experimentally that the nodal lines exhibit sizable dispersion \cite{Y.M. Shao2020,S.Ahn2017,M.B.Schilling2017}, in contrast to an ideal flat line in $k$ space. Its effects have been observed in NLSMs. In particular, the presence of SOC can induce a finite gap $\Delta_{\mathrm{SO}}$ which breaks the nodal lines \cite{C.Q.Li2018}. This will not change the overall topology of the electronic bands, instead it results in a quantitative modification of the Berry phase, $\phi_B=\pm\pi\left(1-\frac{\Delta_{\mathrm{SO}}}{2E_F}\right)$, which depends on both the Fermi energy $E_F$ and the spin-orbit gap $\Delta_{\mathrm{SO}}$. Taking this into account, it is likely that the observed angle-dependent Berry phase reflects the effect of a finite spin-orbit gap.

The Fermi energy can be estimated using the Landau index found in the quantum oscillations as the Dirac bands in NLSMs quantize under magnetic field\cite{Y.M.Shao2019} as $E_{\pm n}=\pm \sqrt{2 e \hbar|n| B \bar{v}^{2} \cos \phi+\Delta^{2}}$, where $\bar{v}$ is the average Fermi velocity normal to the nodal-line direction, $\Delta$ is half of the gap, which in our case is $\frac{\Delta_{\mathrm{SO}}}{2}$, $\phi$ is the angle between the applied magnetic field and the nodal line. For a crude estimate, we use the Fermi velocity $\hbar v_{F}$ of 2.65 eV$\cdot \AA$ along $\Gamma-M$ direction (perpendicular to the nodal line) obtained from a quasiparticle interference measurement\cite{M.S.Lodge2017}. The Fermi energy $E_F$, as counted from the Fermi level to the neutral point, is found to be 167 $\sim$ 173 meV for $0^{\circ} \leq \varphi \leq 45^{\circ}$. $E_F$ varies significantly along the nodal line due to its dispersion. Together with the reported infrared optical \cite{M.B.Schilling2017} and APRES \cite{C.Chen2017} data on \ZSS\ that the $\Delta_{\mathrm{SO}}$ along the nodal lines is 30 $\sim$ 50 meV, the correction to the Berry phase $\phi_{B}=\pm \pi\left(1-\frac{\Delta_{\mathrm{SO}}}{2 E_{F}}\right)$ is found to be as large as 0.15$\pi$, which corresponds to a significant fraction in the observed angular variation of the Berry phase. It is known that the Fermi velocity bears strong suppression along the nodal-line direction, which makes the correction term more important. Earlier quantum oscillation measurements on \ZSS\ mostly focus on the out-of-plane magnetic field configuration \cite{J.A.Voerman2019}. Pressure induced change in the Berry phase were also reported \cite{D.VanGennep2019,C.C.Gu2019}. However, even for the very similar quantum oscillation frequencies, the obtained Berry phase bares significant variation \cite{J.Hu2017,R.Singha2017,M.N. Ali2016,M.Matusiak2017,X.F.Wang2016}. It is likely that in additional to experimental errors, the SOC effect, which was not take into account previously, would play an important role. A more quantitative $k$ dependence of $\Delta_{\mathrm{SO}}$ and $E_F$ is needed to further pin down this possibility.
	
\section{CONCLUSION}

In conclusion, we explored the consequences of the extended nodal line and SOC in quantum transport properties of \ZSS\ and identified distinct features due to the one-dimensional distribution of the nodal line in $k$ space. Our observation of clear field-orientation dependent nontrivial Berry phase points to the importance of SOC and the nodal-line dispersion in the interpretation of quantum transport in NLSMs.

\begin{acknowledgments} Work at SJTU and ZJU is supported by National Key Projects for Research \& Development of China (Grant No. 2019YFA0308602), National Natural Science Foundation of China (Grant No. 11804220, 11974307), and Natural Science Foundation of Shanghai (Grant No. 20ZR1428900). H.X. also acknowledges additional support from a Shanghai talent program. \end{acknowledgments}


\bibliographystyle{apsrev4-1}

\end{document}